\documentclass[manuscript,screen,nonacm]{acmart}
\setcopyright{none}
\acmDOI{}
\acmISBN{}

\usepackage{amssymb}
\usepackage{mathtools}
\usepackage{stmaryrd}
\usepackage{siunitx}
\usepackage{multirow}
\usepackage{array}
\usepackage{verbatim}
\usepackage{enumitem}
\usepackage{subcaption}
\usepackage{xspace}
\usepackage{algorithm}
\usepackage{algorithm}
\usepackage{tabularray}   
\usepackage{pgfplots}     
\usepackage{algpseudocode}

\usepackage[disable]{todonotes}
\newcommand{\gcnote}[1]{\todo[color=red!40]{GC: #1}}

\newcommand{\toolname}{{\sc BaDaaS}\xspace}

\newcommand{\emnistTargetA}{4}
\newcommand{\emnistTargetB}{8}

\newcommand{\emnistTestAcc}{0.9645}
\newcommand{\emnistTestAccA}{0.979}
\newcommand{\emnistTestAccB}{0.950}

\begin{document}

\title{Everyone's a little bit racist: a novel ML backdoor attack in black-box mode}

\author{Michele Cerreta}
\affiliation{
  \institution{IMT School for Advanced Studies}
  \country{Lucca, Italy}}
\email{michele.cerreta@imtlucca.it}
\author{Gabriele Costa}
\affiliation{
  \institution{IMT School for Advanced Studies}
  \country{Lucca, Italy}}
\email{gabriele.costa@imtlucca.it}

\begin{abstract}
Backdoors are among the most impactful attacks against Machine Learning (ML), yet most existing methods assume some form of training-time poisoning or direct model manipulation.
This is often considered sufficient to assume that existing systems are not exposed to actual risks.
In this paper, we introduce Backdoor-as-a-Service, a model-agnostic, fully black-box technique that, instead of installing a backdoor, activates it in the form of a latent secondary-task leveraging on an already present behavior.
Our tool uses genetic search to discover input transformations that induce systematic output differences aligned with an attacker-defined task, without modifying training data, architecture, or weights.
We verify our approach against two targets.
First, we target a MNIST digits classifier that we coerce into implementing a binary letter-recognition task.
Then, we attack a production-grade NSFW classifier using face-specific perturbations to turn it into a proxy for demographic-group classification.
In both cases, our method successfully enables the target backdoor.
Finally, we discuss the existing countermeasure to assess the actual risk for the deployed systems.
\end{abstract}




\maketitle


\section{Introduction}
\label{sec:intro}

Machine learning-based classifiers are often deployed as opaque services, exposing only a limited inference. 
Such a deployment strategy is usually taken to reduce risk in practice.
Indeed, if an attacker cannot influence training data, alter model weights, or inspect internals, then classic poisoning and trojan-installation attacks become much harder to mount. 
This line of reasoning may stem from two implicit assumptions: $(i)$ machine learning models are fragile when exposed to the adversary and $(ii)$ adversarial attacks require access to the model internals (e.g., training data and model weights).
The first statement appears very reasonable, since the model structure and behavior are in a direct relation.
Instead, the second statement is mostly based on the state-of-the-art of adversarial attacks.
To the best of our knowledge, indeed, all the existing approaches fall under this restriction, which amounts to a white- or gray-box working assumption.

Reasonably, among the existing attack scenarios, \emph{backdoors} are among the worst.
In simple terms, a backdoor attack injects a secondary, parasitic task in a model so that the attacker can exploit it for her own business.
For instance, a traffic control system that tracks vehicles might be hacked to carry out an illegal task such as citizen surveillance via face recognition.
In this way, the attacker also abuses the victim system and its resources as a free rider, since the malicious classification task is carried out by the injected model.
Furthermore, the attacker attributes the victim with the responsibility for the malicious behavior.

Nowadays, the main limitation to backdoor attacks is that the existing techniques operate either through data or model poisoning.
Hence, by protecting the internal structure of the models and the training process, system administrators prevent the attacker from installing a backdoor in their classifiers.
In general, it is commonly accepted that a backdoor cannot be activated if the attacker operates in \emph{black-box} mode, i.e., when she can only invoke a deployed model via its classification interface.


In this paper, we present a novel, fully black-box backdoor attack called \emph{Backdoor-as-a-Service}.
The main intuition behind our attack is that latent biases are pervasive in trained models, due to the very inner nature of the machine learning algorithms and their training process~\cite{10.1145/3465416.3483305}.
Thus, the attacker does not need to install a trigger.
Rather, she just has to craft inputs able to exploit the natural behavior of the system.
These inputs can be generated by studying the behavior of a victim classifier \emph{without violating its deployment interface}.

Here we also present \toolname, an open source implementation of our attack.
Rather than installing a backdoor, e.g., during training, \toolname{} searches for an input transformations that make the victim classifier behave according to the attacker-chosen task. 
The transformation is then used to implement a \emph{wrapper} that manipulates the model input and interprets its output.
As a result, the wrapper behaves as a classifier for the secondary task, possibly unrelated to the target model's intended one, which only serves as an internal engine.

The key observation here is that many deployed models expose more information than their advertised task suggests. 
A classifier trained for one purpose may already contain information about other properties of the input distribution, a phenomenon known as overlearning, in which models implicitly learn to recognize attributes outside their training objective~\cite{song2020overlearningrevealssensitiveattributes}. 
If those correlations are stable enough, a black-box adversary can surface them through repeated querying and structured perturbations.
In this sense, \toolname{} does not ``implant'' a backdoor in the conventional poisoning sense.
Instead, it \emph{activates} a latent secondary functionality that is already present in the trained model.

\toolname{} implements this idea through evolutionary search over \emph{perturbation trees}. 
Each tree is a compact transformation program whose leaves apply local brightness, color, frequency, or geometric modifications to spatially partitioned image regions.
The search objective is not simply misclassification of a single input, but differential behavior across the secondary task's output labels. 

We evaluate the approach in two scenarios. 
The first attack, which serves as a working example, targets a handwritten-digit classifier and coerces it into performing binary letter recognition. 
The second attack targets FalconsAI, a production-grade NSFW classifier, and it confirms the practical feasibility of our technique.
In particular, we convert FalconsAI into a classifier of demographic-group membership on face images. 
Our experimental results demonstrate that online classifiers can be exposed to this novel threat.
Furthermore, by considering state-of-the-art countermeasures and mitigations, we discuss the exploitability in the wild.
Our conclusion is that, currently, attackers using this strategy cannot be effectively countered.

Our contribution is threefold.
\begin{enumerate}
  \item We introduce a new black-box, inference-time threat model in which a wrapper can activate a secondary task without retraining or model access.
  \item We design and implement a tool, called \toolname, using genetic algorithms for discovering compact perturbations activating the desired backdoor.
  \item We provide experimental evidence that such perturbations can turn deployed classifiers into unintended secondary-task classifiers even in industrial products.
\end{enumerate}


The remainder of the paper is organized as follows. Section~\ref{sec:background} recalls the concepts used throughout the paper. 
Section~\ref{sec:related} positions our work with respect to backdoors, adversarial reprogramming, and other related attacks. 
Section~\ref{sec:threat} formalizes the attacker model. Section~\ref{sec:badaas} presents the overall \toolname{} architecture. 
Sections~\ref{sec:emnist} and~\ref{sec:nsfw} describe the two attacks carried out with \toolname. 
Section~\ref{sec:results} reports the experimental evaluation and discusses it.
Finally, Section~\ref{sec:conclusion} concludes the paper.

\section{Background}
\label{sec:background}

In this section, we recall some concepts that are fundamental to understanding the rest of the paper.

\subsection{Classification tasks}

In the context of ML, a \emph{classification task} $T$ is the computational problem of \emph{correctly} mapping samples to a finite number of classes.
Each task is described in terms of input and output domains, denoted $I$ and $O$ respectively, which define its signature $T: I \rightarrow O$.
In general, any nontrivial task admits infinitely many implementations, i.e., algorithms respecting the task's signature.
Implementations can be ranked according to their \emph{accuracy}, which is estimated as the ratio of correct predictions to total predictions.
Usually, the ground truth is assumed to be defined \emph{a priori}, e.g., in the form of a labeled training set.

We model a classifier as a mapping $C : I \to \Delta(O)$, where $\Delta(O)$ denotes the probability simplex over $O$.
Given an input $x \in I$, $C(x)$ returns a vector of scores or probabilities $(p_o)_{o \in O}$; a hard label is obtained via $\hat{y} = \arg\max_{o} p_o$.
This probabilistic view is convenient because many modern architectures (e.g., softmax outputs of DNNs) naturally provide confidence scores, which can be exploited by wrappers for aggregation or calibration.

A classical classification task is the \emph{object detection}, in which the goal is to recognize what object appears in the classified image~\cite{liu2020deep}.
For instance, the \emph{handwritten digit recognition problem}~\cite{lecun1998mnist} belongs to this category.
Another example of a classification task is the \emph{content classification} problem, where one aims to establish whether a target image includes certain elements, such as nudity or violence~\cite{arora2023detecting}.
An instance of this class of problems is the \emph{not safe for work} (NSFW) task~\cite{bicho2020nsfw} \cite{Guzman_2023}.


\subsection{Evaluation metrics}

The performance of a classification algorithm is typically measured through a set of evaluation metrics.
Below, we quickly recall them.

Accuracy measures how often the algorithm succeeds in its task. 
It is computed as the rate of successful experiments, in symbols $T/(T+F)$, where $T$ and $F$ are the number of true and false results, respectively.
Often, in a classification task, $T$ and $F$ are further split into positive and negative, i.e., $T = TP + TN$ and $F = FP + FN$.
Accuracy alone is often insufficient to characterize the classifier behavior, especially under distributional shifts or when classes are imbalanced.
We therefore employ other metrics.

A confusion matrix $M \in \mathbb{N}^{n \times n}$ for a $n$-class task counts the occurrences of actual/predicted pairs, exposing class-wise error modes.
Precision and recall measure the performance of the classifier with respect to the positive, in symbols $TP/(TP+FP)$, and true, in symbols $TP/(TP+FN)$, classes.
Finally, F1 score combines precision and recall to evaluate the efficacy of the classifier.
In symbols, 
\[
\textnormal{F1} = 2 \cdot \frac{\textnormal{Precision} \cdot \textnormal{Recall}}{\textnormal{Precision} + \textnormal{Recall}}.
\]

\section{Related work}
\label{sec:related}

In this section we survey on the work on backdoor and other related attacks.

\subsection{Backdoor attacks}

In machine learning, backdoor (aka trojan) attacks aim to embed hidden functionality into an existing model.
The goal is to force a desired behavior on certain, malicious inputs, possibly while the system behaves normally on benign inputs~\cite{9743317,chen2017targetedbackdoorattacksdeep,Saha_Subramanya_Pirsiavash_2020}.
The behavioral change is activated through some sort of \emph{trigger} that the victim model recognized. 
Early surveys discussing backdoor attacks in deep neural networks (DNNs) for vision tasks and beyond are, e.g.,~\cite{9743317,gao2020backdoorattackscountermeasuresdeep}.
Known trigger injection strategies typically rely on one of two approaches, i.e., poisoning of the training dataset and direct manipulation of model weights or architecture.
In the first case, the attacker modifies the training data so that the victim model is also trained for the secondary, parasitic task.
In the second one, the attacker enables the malicious task by tampering with the internal logic of the model.
BadNets~\cite{gu2019badnetsidentifyingvulnerabilitiesmachine} is a canonical instance of the former, embedding patch triggers in poisoned samples; Trojaning Attack~\cite{Trojannn} exemplifies the latter via neuron-targeted weight retuning.

In~\cite{schwarzschild2020toxic}, Schwarzschild~et~al. compare a wide range of poisoning and backdoor attacks under standardized settings.
There, the authors show that the effectiveness of these techniques varies greatly with the victim's architecture and training hyperparameters.
Among the most sophisticated poisoning attacks, MetaPoison~\cite{metapoison2020} scales clean-label data poisoning to realistic architectures and even targets hosted AutoML services.
Even in this advanced setting, the attacker still requires effective control over the training pipeline, a capability our threat model explicitly excludes.

In the visual domain, backdoor triggers may be simple patches (e.g., a colored sticker in one corner), blended patterns, semantic triggers (an object overlay), or even input-aware triggers that adapt to the input content.
Saha~et~al.~\cite{Saha_Subramanya_Pirsiavash_2020} introduce hidden-trigger backdoors, where the trigger is absent from the poisoned samples and surfaces only at inference.
From a defensive angle, Zeng~et~al.~\cite{zeng2021frequency} show how many classic triggers include high-frequency artifacts that can facilitate detection.

The vast majority of these attacks assume some form of backdoor \emph{installation}, via training-set poisoning, weight tampering, or architectural modification, and require, at minimum, control over the training pipeline.
Dynamic or input-aware variants~\cite{9797338} adapt the trigger to the input but still depend on training-time installation.
\toolname{} operates under a strictly disjoint set of assumptions: no training data is altered, no weight is modified, and no auxiliary training is performed.
The trigger is a deterministic input transformation applied at inference time, and the wrapper activates decision biases already embedded in a deployed model.
This positions \toolname{} as an instance of \emph{inference-only, triggerless} backdoor activation, distinct from both static- and dynamic-trigger families in threat model and in the defenses they require.

\subsection{Adversarial reprogramming}
\label{sec:related:reprogramming}

Elsayed~et~al.~\cite{elsayed2018adversarial} introduced \emph{adversarial reprogramming}, demonstrating that a pre-trained ImageNet classifier can be repurposed to perform an entirely different task (e.g., MNIST digit classification) by learning a single universal adversarial perturbation added to all inputs.
Subsequent work has explored reprogramming for fraud detection~\cite{10.1145/3459637.3482053}, text classification~\cite{neekhara2018adversaria}, and other domains, showing broad applicability.
Zheng~et~al.~\cite{zheng2023whyadversarial} provide theoretical analysis of when and why adversarial reprogramming succeeds, showing that success depends on input gradient alignment and dimensionality.


Adversarial reprogramming and \toolname{} share the high-level goal of inducing a model to perform a task it was not trained for, but differ in three substantive ways.
First, reprogramming typically assumes white-box or query-with-gradient access to optimize a single universal additive perturbation; \toolname{} assumes strictly black-box access.
Second, reprogramming hijacks general model capacity through a dense pixel-level perturbation, whereas \toolname{} evolves compact, interpretable transformation programs and explicitly targets latent decision biases rather than raw capacity.
Third, the two methods consequently demand different defenses: input sanitization disrupts reprogramming but does not address the bias substrate that \toolname{} exploits.

\subsection{Evolutionary search and programmatic perturbations}

Beyond triggers, another relevant axis is the generation of adversarial or malicious perturbations via gradient-free or evolutionary search.
The study of adversarial examples originated with the seminal observation by Szegedy~et~al.~\cite{szegedy2013intriguing} that imperceptible perturbations can cause confident misclassifications in deep networks.
Traditional adversarial examples rely on gradient information, either directly in white-box settings or by transferring perturbations crafted on a substitute model trained to mimic the target~\cite{papernot2017practical}; in contrast, evolutionary or programmatic perturbations operate where gradients are unavailable, models are black-box, or the search space is highly structured.
For adversarial generation specifically, Mercuri~et~al.\ describe evolutionary techniques that craft adversarial examples using semantic transformations and fitness guided by the target network's output~\cite{mercuri2023evolutionary}. Particularly close in spirit, POBA-GA~\cite{chen2019pobaga} uses a black-box genetic algorithm to optimize perturbations under a probability-margin objective with an explicit perturbation penalty.

Programmatic perturbations (e.g., tree-structured color masks or spatial transformations) extend this line of work by encoding compact transformation programs rather than raw pixel perturbations.
Compared to pixel-level adversarial noise, programmatic approaches produce human-interpretable patterns and often include size or complexity penalties to favor simplicity.
For instance, Bradley~and~Blossom\ apply genetic algorithms to produce adversarial examples that remain visually credible and semantically consistent with the input~\cite{bradley2023visuallycredible}.
\toolname{} adopts the probability-margin objective with explicit perturbation penalty introduced by POBA-GA~\cite{chen2019pobaga}, but repurposes it for a different goal: whereas POBA-GA and related evolutionary adversarial methods pursue evasion or misclassification of \emph{individual} inputs, \toolname{} exploits population-level asymmetries to transform a biased classifier into a secondary-task activator.
This coupling of programmatic genetic search with bias exploitation is, to our knowledge, distinct from both adversarial-example generation and conventional backdoor attacks.

\subsection{Fairness attacks and bias exploitation}
\label{sec:related:fairness}

Buolamwini and Gebru~\cite{pmlr-v81-buolamwini18a} document systematic accuracy disparities across demographic groups in commercial gender classifiers, evidence that latent demographic correlations are pervasive in production models.
A growing body of work then studies attacks that \emph{exacerbate} rather than mitigate this bias.
Mehrabi~et~al.~\cite{mehrabi2021exacerbating} demonstrate poisoning attacks that increase demographic disparity in classifiers trained with fairness constraints.
Chang~et~al.~\cite{chang2020adversarial} show that fair machine learning algorithms can be manipulated via adversarial sampling to produce biased outcomes on held-out data.
In the face-attribute domain, Zeng~et~al.~\cite{zeng2023adversarial} study the adversarial robustness of demographic fairness, showing how malicious perturbations can degrade fairness in face attribute recognition, and proposing fairness-aware robust training as a defense.
More broadly, Mehrabi~et~al.~\cite{10.1145/3457607} provide a comprehensive survey of bias and fairness in machine learning, cataloguing sources of bias and mitigation strategies.

These fairness attacks operate at \emph{training time}, poisoning data or manipulating the training process to amplify bias.
\toolname{}, by contrast, assumes the model is already deployed and exploits \emph{existing} latent biases at \emph{inference time} without any training-phase intervention.
Our work thus complements the fairness-attack literature by showing that demographic biases can be weaponized post hoc, even when training data and model weights are inaccessible.
This inference-time exploitation represents a distinct threat model that current fairness defenses do not address.

\section{Attacker model}
\label{sec:threat}

In this section, we describe the reference attacker model.
As for any backdoor attack, the goal is to implement a secondary task $T' : I' \rightarrow O'$ by leveraging an existing classifier $M$, well-trained for a compatible, yet unrelated task $T : I \rightarrow O$.
However, the distinguishing feature of our attack is that the attacker has absolutely no control over the victim classifier.
In particular, we assume that the attacker:
\begin{enumerate}
    \item \emph{cannot} tamper with the training process of $M$;
    \item \emph{cannot} access the internal structure and state of $M$; 
    \item \emph{cannot} observe the behavior of $M$ when used by others;
    \item \emph{can} craft inputs, submit them to $M$ and check the output.
\end{enumerate}
All in all, these assumptions sum up to a full black-box attacker model, where the adversary can only interact with the target system according to its legal, I/O interface.

Under our attacker model, the attacker \emph{cannot} install a backdoor using techniques that require model manipulation, e.g., through data or model poisoning.
Instead, the backdoor must be \emph{activated} by means of an existing bias in $M$.
To achieve that, the attacker can only develop a \emph{wrapper} $W$, i.e., an infrastructure that $(i)$ crafts the inputs and $(ii)$ interprets the outputs of $C$.
In other terms, $W$ computes $T'$ by giving it a $T$-interpretation that $M$ will process, thus acting as a \emph{confused deputy}~\cite{hardy1988confused}.

\begin{figure}
    \centering
    \includegraphics[width=\columnwidth]{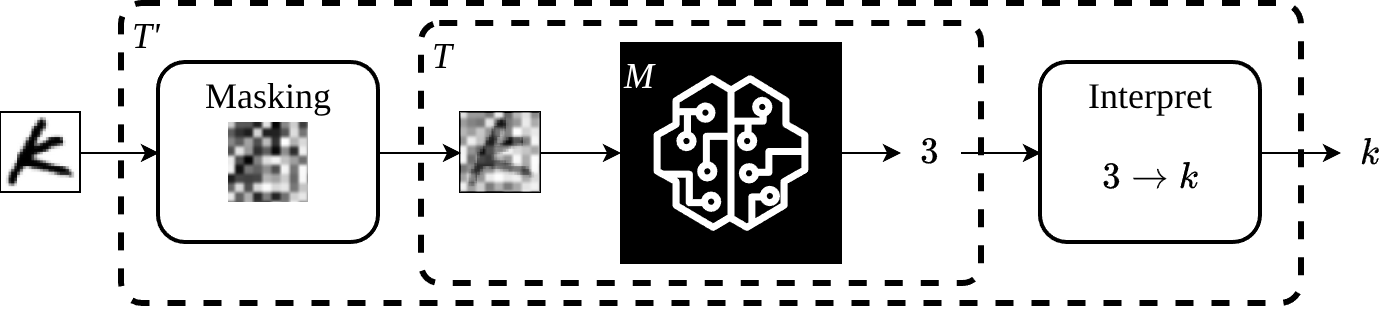}
    \caption{Overall schema of the reference attacker model.}
    \label{fig:threat}
\end{figure}

The overall attacker model is schematized in Figure~\ref{fig:threat}.
There, we exemplify the attack scenario for a classical digit classification task $T$ (inner, dashed box).
Instead, the secondary task $T'$ consists of a single letter discrimination.
The attacker's goal is to obtain an accurate classifier for $T'$ using the wrapped/attacked model. 
Reasonably, $T'$ cannot be more complex than $T$, since $M$ might lack the resources to carry out the desired task.
For instance, if $M$ can accurately classify 10 digits, it might not be adequate to distinguish between all 26 letters, while the opposite may be doable.
In symbols, this amounts to requiring $|I'| \leq |I|$ and $|O'| \leq |O|$.
Hence, in the following, we will often consider $T'$ as a binary classification task.

 Another aspect of interest is that of the wrapping strategy.
 In general, any translation from $I'$ to $I$ and interpretation from $O$ to $O'$ may be considered.
 For the sake of presentation, here we restrict to image classification tasks for both $T$ and $T'$.
 Even under this assumption, there might be multiple translation strategies.
 Among them, we focus on mask overlays.
 The main reason is that masks can be generated using efficient fuzzing algorithms, as the one given in Algorithm~\ref{alg:evolve}.
 In this way, the attacker also has an economic advantage with respect to, e.g., adversarial training.

\section{Introducing \toolname{}}
\label{sec:badaas}

Below, we present \toolname, an open source\footnote{The source code and the experimental artifacts will be shared in case of paper acceptance.} tool implementing the attack scenario described in Section~\ref{sec:threat}.
Briefly, \toolname{} automatically generates the wrapping components needed to transform a target image classifier ($T$) into an arbitrary, binary image classifier ($T'$).


\begin{figure*}
    \centering
    \includegraphics[width=0.98\textwidth]{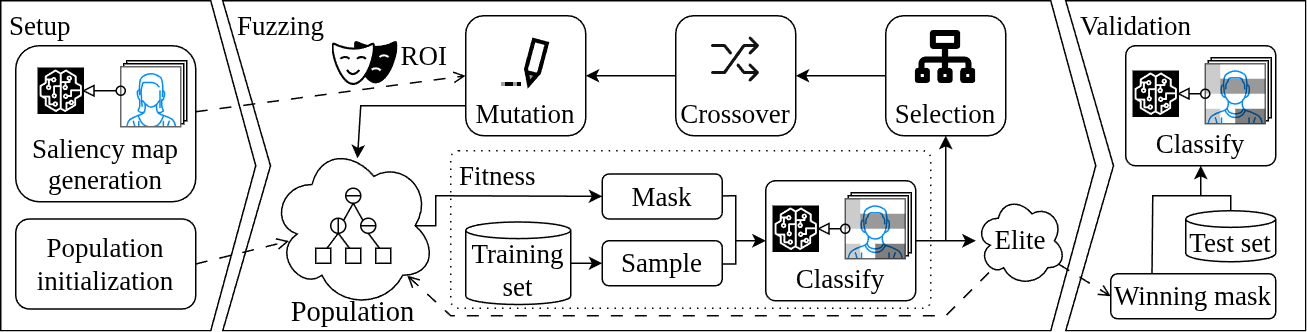}
    \caption{The general workflow of \toolname.}
    \label{fig:workflow}
\end{figure*}

\subsection{Overall workflow}
\label{sec:badaas:arch}

\toolname follows the three-stage workflow depicted in Figure~\ref{fig:workflow}.
Briefly, the three stages are the following.

\begin{description}
    \item[Setup.] In this phase \toolname executes the preparatory steps for instructing the fuzzing process.
    In particular, two operations are executed, i.e., the \emph{Saliency map generation} and the \emph{population initialization}.
    Saliency maps are constructed using the algorithm presented in \cite{simonyan2014deep}.
    More precisely, saliency maps are used by the mutation operator to identify the \emph{region of interest} (ROI), i.e., locations having the highest relevance for the target classifier.
    This aspect is described in detail in Section~\ref{sec:roi}.
    Instead, the population initialization creates the first generation of random individuals.
    Individuals encode masks to be applied to the target images.
    The encoding occurs through a \emph{mask encoding tree} (MET) and the initialization relies on a few hyperparameters defining, e.g., the initial MET depth.
    This aspect is discussed in Section~\ref{sec:met}.
    \item[Fuzzing.] The fuzzing phase is implemented through the procedure in Algorithm~\ref{alg:evolve}.
    Most of the elements composing the fuzzing algorithm are standard but for a few exceptions.
    In particular, the algorithm $(i)$ uses a saliency-based mutation operator (see Section~\ref{sec:roi}), $(ii)$ evolves a population of MET (see Section~\ref{sec:met}), and $(iii)$ interacts with the victim classifier to compute the fitness of individuals (see Section \ref{sec:fitness}).
    Furthermore, the fuzzing process uses elitism \cite{10.1007/978-3-540-71805-5_69}, i.e., it ensures fitness monotonicity by preventing high-fitness individuals from being discharged in the next generation.
    In Algorithm~\ref{alg:evolve}, the elite $E$ is extracted (line 5), added to the selection output during crossover and mutation (line 7) and restored in the next generation (line 8).
    The procedure terminates when one of three conditions holds: a target accuracy is reached (\textsc{GoalReached}), the best fitness fails to improve over a fixed number of generations (\textsc{Stagnation}), or the maximum number of generations is exceeded (\textsc{MaxGenerations}).
    \item[Validation.] When the fuzzing process terminates, the winning mask is taken from the elite group.
    Using a separate test set for task $T'$, the validation phase interacts with the victim classifier and measures the accuracy of the obtained mask.
\end{description}

\renewcommand{\algorithmicrequire}{\textbf{Input:}}
\renewcommand{\algorithmicensure}{\textbf{Output:}}

\begin{algorithm}[t]
\caption{Genetic fuzzing mask generation algorithm}
\label{alg:evolve}

\begin{algorithmic}[1]
\Statex \textbf{Input:} Samples $I'$; Labeling function $L' : I' \rightarrow O'$;
\Statex \textbf{Output:} Mask $m$;
\Statex \textbf{Hyperparameter:} Initial depth $\delta$; Window size $\omega$; Slide factor $\varphi$; Sliding rate $\rho$; Population size $\pi$; 
\Statex $\qquad$ Tournament size $\tau$; Mutation rate $\alpha$; Crossover rate $\gamma$; Elite size $\varepsilon$;

\State $P \gets \textsc{InitPop}(\pi, \delta);$ \Comment{Generate initial population}
\State $I'_B \gets \textsc{Sample}(I', L', \omega);$
\Comment{Initial balanced sampling}
\Repeat
  \State $f \gets \textsc{Fitness}(P, I'_B);$ \Comment{Evaluate population}
  \State $E \gets \textnormal{arg top}_{\varepsilon,f} P;$ \Comment{Save elite individuals}
  \State $W \gets \textsc{Select}(P, f, \tau);$ \Comment{Tournament selection}
  \State $P' \gets \textsc{Crossover\&Mutate}(W \cup E, \gamma, \alpha, \pi - \varepsilon);$
  \State $P \gets P' \cup E;$ \Comment{Restoring elite in next generation}
  
  \If{$\textsc{CurrentGen} \equiv_\rho 0$}
    \State $I'_B \gets \textsc{Slide}(I'_B, I', L', \varphi);$ \Comment{Slide window}
  \EndIf
 
\Until{GoalReached $\lor$ Stagnation $\lor$ MaxGenerations}
\State $t^* \gets \textnormal{arg top}_{1,f} P;$ \Comment{Select winner MET}
\State $m \gets\textsc{MaskFromTree}(t^*);$ \Comment{Convert $t^*$ to mask}
\end{algorithmic}
\end{algorithm}

\subsection{Mask encoding trees}
\label{sec:met}

A \emph{mask encoding tree} (MET) is the data structure used by \toolname to encode and modify image masks.
Briefly, a MET is a binary tree in which internal nodes represent one of two split modalities, i.e., vertical $\varominus$ and horizontal $\varobar$.
Intuitively, the two child nodes of $\varominus$ define the upper and lower side (respectively) of a mask, while the children of $\varobar$ define the left and right side (respectively).
Instead, MET leaves ($\square$) contain a differential operator that modifies the area of the picture it applies to.
Differential operators may vary with the image pixels encoding.
For instance, for one-byte black and white pixels, the differential operator may amount to a single signed byte.

\begin{figure}[t]
    \centering
    \includegraphics[width=0.98\linewidth]{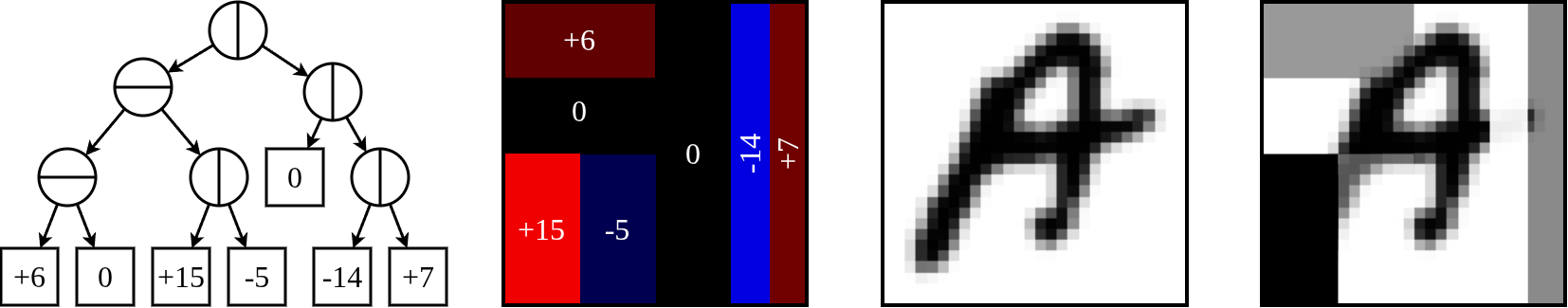}
    \caption{An example MET (left), its corresponding mask (center left), and its application (right) to a sample (center right).}
    \label{fig:mask}
\end{figure}
\gcnote{ora la figura dell'albero è troppo grande, proverei a ridurla. Fatto, guarda se ti aggrada}

Figure~\ref{fig:mask} shows an example MET and its application to a picture.
Briefly, the tree root splits the mask horizontally.
The left side is then split vertically and its upper and lower submasks are then split vertically and horizontally (respectively).
Instead, the root's right child splits its submask horizontally.
The left-hand side is a leaf node, while the right-hand one is a further horizontal split.
Leaves contain differential operators for a four-bit pixel encoding, where 0 stands for white and F (15) stands for black.
Thus, positive leaf values will result in darkening regions, while negative ones will make the target region brighter. 
In the bottom part of Figure~\ref{fig:mask}, we show the resulting mask using false colors, red for darkening and blue for brightening regions (respectively).
When applying the mask $m$ to a sample $x$, in symbols $m \otimes x$, the result is the one shown on the right, where pixels are recolored depending on the differential operators.

The maximum depth of a MET can be computed as the logarithm of the target image dimension.
For instance, if we consider a $8 \times 8 = 64$ pixel image, a depth 6 binary tree is sufficient to encode pixel level transformations.
The same holds for the differential operators encoded by the leaves.
As a consequence, MET are general enough to encode any possible image transformations.
Nevertheless, large trees are undesirable for several reasons and, primarily, for the efficiency of the fuzzing process.
Hence, \toolname will tend to privilege smaller MET.
This aspect is further detailed in Section~\ref{sec:fitness}.

\subsection{Saliency-based mutation}
\label{sec:roi}

\toolname{} uses a saliency-guided mutation operator to direct the search procedure.
The reason is that, as discussed above, the search space of the fuzzing algorithm is extremely large.
In general, assuming squared pictures of size $h \times w$, the number of possible masks is $h \cdot w \cdot p$, where $p$ is the dimension of the encoding of each pixel.
For instance, $128 \times 128$, RGB pictures where each color channel is encoded as a single byte result in more than $2.7 \cdot 10^{11}$ possible masks.
Thus, an efficient mutation strategy is fundamental to avoid pure random search.

Intuitively, the saliency-based mutation works by mutating more salient regions of a MET with higher probability.
In practice, a saliency map is generated using the algorithm described in \cite{simonyan2014deep}.
In general, given an input picture, a saliency map $s \in [0,1]^{h \times w}$ assigns to each pixel $p = (p_x,p_y)$ a score $s(p_x,p_y)$ (or simply $s(p)$) measuring the relevance of the pixel for the classification of the picture.
Given a set $D$ of samples, we can compute the overall saliency map as the average of the saliency maps of each $d \in D$.

Since our goal is to hijack the classification process by reusing the original output labels, we compute a \emph{differential} saliency map, which we call \emph{region of interest} (ROI).
To do that, we compute the difference between the two highest saliency maps among $s_1, \ldots, s_n$ (i.e., one for each label in $O$).
In symbols, we have the following.

\begin{equation}
s(p) = \max_{i \in \{1,\ldots,n\}} s_i(p) - \max_{j \in \{1,\ldots,n\}\setminus \{i\}} s_j(p)
\end{equation}

When mutating a MET $t$, we compute all the image regions $R_\ell$ denoted by each leaf $\ell$ of $t$.
For each $R_\ell$, we compute its overall saliency score $\bar{s}_\ell$ as the average of the ROI saliency of the pixels belonging to $R_\ell$.
In symbols,

\begin{equation}
\bar{s}_\ell = \frac{1}{|R_\ell|} \sum_{p\in R_\ell} s(p).
\end{equation}

When a mutation occurs, the operator selects one leaf with probability
\begin{equation}
P(\ell) =
\frac{\bar{s}_\ell^{a} + c}
{\sum_{\ell'} \left(\bar{s}_{\ell'}^{a} + c\right)},
\end{equation}
where the hyperparameter $a \geq 1$ is used to amplify the relative probability of the high saliency regions (by default \toolname sets $a=3$), and $c > 0$ preserves a non-zero exploration probability for all regions, regardless of their saliency.

The mutation operator then modifies the selected leaf by randomly picking one of three possible mutation actions, i.e., jittering, expansion, or contraction.
Briefly, jittering modifies the differential operator of the leaf, expansion replaces the leaf with a randomly generate sub-tree and contraction replaces the sub-tree containing the leaf with the leaf itself.
Also, to avoid MET size explosion, a maximum depth is set through a hyperparameter and expansion is disabled when it would cause exceeding it.

\subsection{Fitness function}
\label{sec:fitness}

\toolname{} evaluates each MET by measuring how accurately the victim model carries out the backdoor task $T'$ when submitting masked images.
To this aim, a training dataset for $T'$ is used.
Although a large dataset is crucial to drive the masks' evolution, testing each MET against each element of the dataset would quickly become impractical.
For this reason, \toolname resorts to sampling, i.e., a smaller, randomly selected subset of the training data is used to evaluate the masks (see Algorithm~\ref{alg:evolve}, line 2).
Periodically, the training set is resampled to avoid overfitting. 
A training set for a $n$-ary classification task has the form $\mathcal{D} = \{D_1, \ldots D_n\}$ where each $D_i$ contains samples of the $i$-th label. 

In detail, the fitness function is computed as follows.
Depending on the configuration of the victim model $M$, the attacker defines an evaluation function $\phi_M : O \rightarrow \mathbb{R}$ for $M$'s response.
For instance, $\phi_M$ can return the confidence score that $M$ assigns to the generated output label. 
Thus, given a mask $m$ and a sample $x$, the attacker computes $\phi_M(m \otimes x)$ to test $M$.

Then, given a MET $t$ encoding mask $m$, we compute the accuracy vector $\mu(t)$ as follows.
\begin{equation}
\label{eq:accvec}
\mu(t) = \left(\sum_{x \in D_1} \frac{\phi_M(m \otimes x)}{|D_1|}, \ldots, \sum_{x \in D_n} \frac{\phi_M(m \otimes x)}{|D_n|}\right)
\end{equation}
Eventually, the accuracy vector $\mu(t)$ is used for calculating the fitness function.
\begin{equation}
\label{eq:fitness}
\mathit{fitness}(t) = \Vert \mu(t)\Vert - \beta Z(t),
\end{equation}

where $\Vert \cdot \Vert$ is the vectorial norm and $Z(t)$ is a perturbation penalty term (scaled by a control parameter $\beta > 0$) as defined in~\cite{chen2019pobaga}.
The perturbation penalty $Z(t)$ is meant to favor smaller MET over large ones.
For instance, \toolname computes $Z(t)$ in terms of the depth of $t$ as follows.
\begin{equation}
    Z(t) = \frac{depth(t)}{\lceil\log_2 (h \cdot w)\rceil}
\end{equation}

\section{\toolname demonstration}
\label{sec:emnist}

The goal of this section is to demonstrate \toolname applied to a working example.
In particular, here we target a CNN trained for the classical handwritten digit recognition ($T$).
The secondary task $T'$ is that of distinguishing between two attacker-chosen, handwritten letters.

\subsection{General setup}
\label{sec:emnist:setup}

\paragraph*{Target classifier}
The victim classifier is a CNN trained on MNIST digits dataset~\cite{lecun1998mnist} till achieving accuracy $>99\%$.
The target architecture consists of two convolutional layers (32 and 64 filters, $3 \times 3$ kernels) with max-pooling, followed by a 128-unit dense layer with dropout and a 10-class softmax output.
The model is treated as a black-box oracle only returning labels in $O = \{0, \ldots, 9\}$.

\paragraph*{$T'$ dataset}
We start from the EMNIST handwritten letters dataset~\cite{cohen2017emnist}.
As for MNIST, samples are $28 \times 28$ grayscale images.
For this experiment, we set $O' = \{A,B\}$, i.e., the attacker aims at implementing a binary classifier distinguishing between characters $A$ and $B$.  
The resulting binary A/B dataset is derived directly from EMNIST and evaluated with a disjoint holdout split (90\% for training and 10\% for the final evaluation).


\subsection{Mask discovery}
\label{sec:emnist:discovery}

We generate masks as described in Section~\ref{sec:badaas}.
For this attack, we only consider \emph{brightness perturbations}, as MNIST/EMNIST images are grayscale.
Also, we assume that the attacker uses labels \emnistTargetA{} and \emnistTargetB{} to classify A and B, respectively.

\paragraph*{Fitness function}
Since the attacker is implementing a binary classifier, Equation \ref{eq:accvec} becomes
\[
\mu(t) = \left(\sum_{x \in D_A} \frac{\phi_{\emnistTargetA}(m \otimes x)}{|D_A|}, \sum_{x \in D_B} \frac{\phi_{\emnistTargetB}(m \otimes x)}{|D_B|}\right)
\]
where $\phi_{\emnistTargetA}(x)$ and $\phi_{\emnistTargetB}(x)$ return the confidence score of the model for sample $x$ w.r.t. label \emnistTargetA{} and \emnistTargetB{}, respectively.
Finally, we set $\beta = 0.02$ and $Z(t) = \mathit{depth}(t)/10$ (since $\log_2(28^2) \simeq 9.6$); Equation~\ref{eq:fitness} therefore becomes
\[
\mathit{fitness}(t) = \Vert \mu(t) \Vert - 0.02 \cdot \frac{\mathit{depth}(t)}{10}.
\]

\paragraph*{Hyperparameters}

For this experiment, we assign the following values to the hyperparameters of Algorithm~\ref{alg:evolve}.

\begin{center}
\begin{tabular}{ccccccccc}
    $\delta$ & $\omega$ & $\varphi$ & $\rho$ & $\pi$ & $\tau$ & $\alpha$ & $\gamma$ & $\varepsilon$  \\
    \hline
    7 & 400 & 0.2 & 50 & 32 & 3 & 0.09 & 0.70 & 4 \\
\end{tabular}    
\end{center}

Evolutionary algorithm parameters $\tau$, $\alpha$, $\gamma$ and $\varepsilon$ are straightforward.
\toolname initializes the population with 32 ($\pi$) random MET of depth 7 ($\delta$).
Training datasets $D_A$ and $D_B$ both contain 400 samples $\omega$ randomly selected from EMNIST restricted to A and B, respectively. 
Every 50 rounds ($\rho$) 20\% ($\varphi$) of the training datasets is resampled.
Finally, we set the stopping conditions to 400 maximum generations.





\subsection{Results summary}
\label{sec:emnist:results}

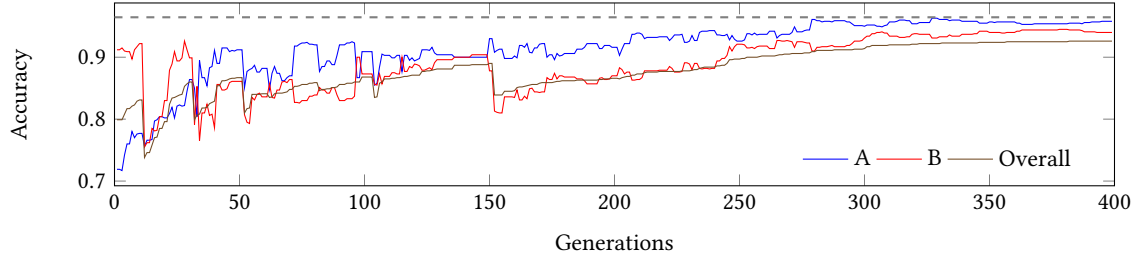
\begin{figure}[t]
\begin{tikzpicture}
\begin{axis}[
    width=0.98\textwidth,
    height=4cm,
    xlabel={Generations},
    ylabel={Accuracy},
    xmin=0,
    xmax=400,
    legend pos=south east,  
    legend columns=3,      
    legend image post style={mark=none},
    legend style={
        draw=none,        
        fill=none       
    }
]
\addplot table [x=gen, y=A, col sep=comma, mark=none] {data/emnist.csv};
\addlegendentry{A}
\addplot table [x=gen, y=B, col sep=comma, mark=none] {data/emnist.csv};
\addlegendentry{B}
\addplot table [x=gen, y=acc, col sep=comma, mark=none] {data/emnist.csv};
\addlegendentry{Overall}
\draw[dashed, thick, gray] 
    ({rel axis cs:0,0}|-{axis cs:0,0.9645}) -- 
    ({rel axis cs:1,0}|-{axis cs:0,0.9645});
\end{axis}
\end{tikzpicture}
\caption{Evolution of the best mask for the EMNIST.}
\label{fig:emnistplot}
\end{figure}

\begin{figure}[t]
    \centering
    \includegraphics[width=\columnwidth]{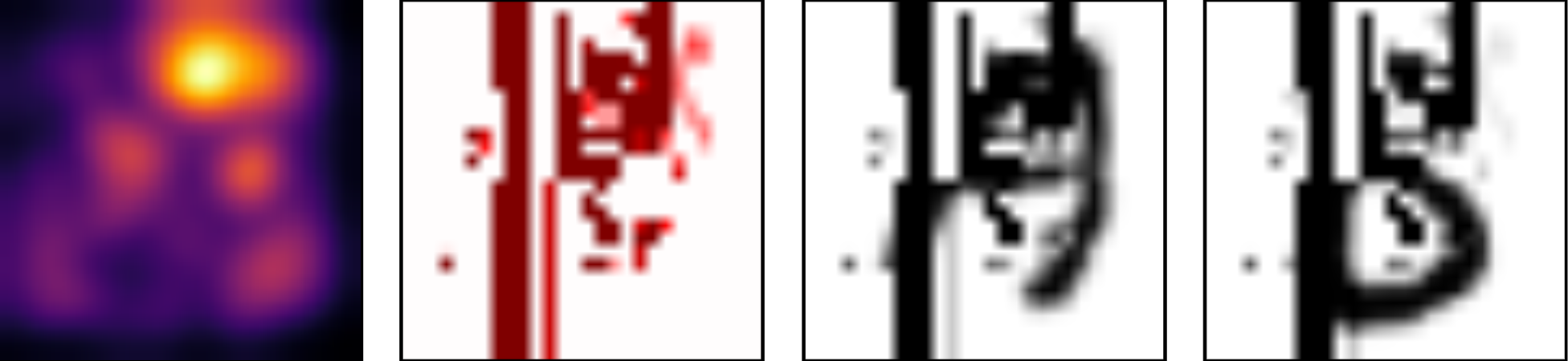}
    \caption{From left to right, ROI, winning mask and applications to two samples for the EMNIST experiment.}
    \label{fig:emnist-result}
\end{figure}

Figure~\ref{fig:emnistplot} shows the per-class and overall accuracy of the best individual at each generation.
The final training accuracy is $92.6\%$.
The evaluation set $D=D_A\cup D_B$, such that $|D_A| = |D_B| = 1000$, consists of 2000 randomly selected samples. On this set, the winning mask reaches \emnistTestAcc{} balanced accuracy (Figure~\ref{fig:emnistplot}, dashed line), with $A=\emnistTestAccA{}$ and $B=\emnistTestAccB{}$ per class. Notably, this number exceeds the search-time peak of $92.6\%$, which confirms that the exported mask generalizes to held-out samples rather than overfitting the search subset.
Figure~\ref{fig:emnist-result} shows some artifacts generated by \toolname, including the winning mask and its application to two samples.

In our view, the EMNIST setting is markedly easier than the FalconsAI one of Section~\ref{sec:results}. Indeed, here the secondary task is close to the victim's training task (digit-label routing rather than demographic prediction on top of an unrelated NSFW classifier), the input domain is small ($28 \times 28$ grayscale), and a single brightness primitive is sufficient.

\section{\toolname in the wild}
\label{sec:nsfw}

In this section, we demonstrate \toolname by activating a backdoor in a state-of-the-art industrial image classifier for the \emph{Not Safe For Work} (NSFW) task~\cite{bicho2020nsfw}.
Briefly, NSFW aims at distinguishing between unsafe images, e.g., those containing sex and violence, and safe ones.


\subsection{General setup}
\label{sec:nsfw:setup}

\paragraph*{Target Classifier}
We consider FalconsAI~\cite{huggingfaceFalconsAi26} state-of-the-art, industrial NSFW classifier.
The classifier is based on a ViT-base-patch16-224 model~\cite{dosovitskiy2021imageworth16x16words} served via the HuggingFace \texttt{transformers} library. 
The model outputs a softmax~\cite{10.1007/978-3-642-76153-9_28, goodfellow2016deep} probability over the two classes \texttt{normal} and \texttt{nsfw}.
In particular, $O = [0,1]^2$ and the meaning of output $(o_1, o_2)$ is that the input image is classified as \texttt{normal} when $o_1 > o_2$ and, vice versa, \texttt{nsfw} when $o_1 \leq o_2$.

\paragraph*{$T'$ dataset}
We use the UTKFace~\cite{zhang2017utkface} dataset, which consists of over 20,000 face images.
Each face is labeled with age, from 0 to 116, gender, male (0) and female (1), and ethnicity, between white (0), black (1), asian (2), indian (3), and other (4).
Our experiments focus on binary classification tasks on these labels.
In particular, we consider the following three tasks.
\begin{enumerate}
    \item $O'_1 = \{\textnormal{black},\textnormal{white}\}$
    \item $O'_2 = \{\textnormal{age} \leq 25, \textnormal{age} > 25\}$
    \item $O'_3 = \{\textnormal{male}, \textnormal{female}\}$
\end{enumerate}
Due to the specificity of the tasks, we also reduced the UTKFace dataset by removing irrelevant samples (e.g., for $O'_1$, we do not consider ethnicity labels other than 0 and 1).
Thus, to keep the two classes equally distributed, we reduced the used dataset to 8000 samples, split into 7000 ($\sim$ 87.5\%) for training and 1000 ($\sim$ 12.5\%) for validation.



\subsection{Mask discovery}
\label{sec:nsfw:discovery}

We evolve masks using the genetic-programming framework of Section~\ref{sec:badaas}, with the following settings.

Leaves perturbations belong to these five categories.
\begin{itemize}
  \item Brightness: a random scale of all RGB channels.
  \item Channel: a random scale of a single RGB channel.
  \item Alpha: a random RGBA overlay color.
  \item Compression: a random DCT-based compression~\cite{wallace1992jpeg}.
  \item Distortion: a random grid-based geometric distortion.
\end{itemize}

\begin{figure}[t]
    \centering
    \includegraphics[width=\columnwidth]{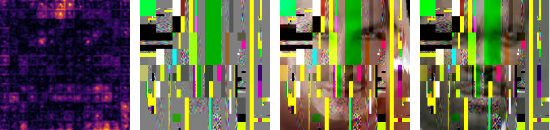}
    \caption{From left to right, ROI, winning mask and applications to two samples for the FalconsAI experiment.}
    \label{fig:roi_mask_samples_falconsai}
\end{figure}



\paragraph*{Fitness function}
Similarly to the EMNIST scenario, Equation~\ref{eq:accvec} becomes
\[
\mu(t) = \left(\sum_{x \in D_A} \frac{\phi_{1}(m \otimes x)}{|D_A|}, \sum_{x \in D_B} \frac{\phi_{2}(m \otimes x)}{|D_B|}\right)
\]
where $\phi_{1}(x)$ and $\phi_{2}(x)$ return the softmax probability of the model for sample $x$ w.r.t. label \texttt{normal} and \texttt{nsfw}, respectively (see Section~\ref{sec:nsfw:setup}).
Also, we set $\beta = 0.02$ and $Z(t) = \mathit{depth}(t)/16$ (since $\log_2(224^2) \simeq 15.6$), which results in Equation~\ref{eq:fitness} becoming
\[
\mathit{fitness}(t) = \Vert \mu(t) \Vert - 0.02 \cdot \frac{\mathit{depth}(t)}{16}.
\]

\paragraph*{Hyperparameters}
In this case, we assign the following values to the hyperparameters of Algorithm~\ref{alg:evolve}.

\gcnote{visto che c'è spazio in orizzontale potremmo mettere le etichette per esteso nella tabella. Fatto}
\begin{center}
\begin{tabular}{ccccccccc}
    Init. depth & Win. size & Slide fact. & Slide rate & Pop. size & Tour. size & Mut. rate & Cross. rate & Elite size \\
    \hline
    $\delta = 7$ & $\omega = 400$ & $\varphi = 0.2$ & $\rho = 50$ & $\pi = 32$ & $\tau = 3$ & $\alpha = 0.09$ & $\gamma = 0.70$ & $\varepsilon = 4$ \\
\end{tabular}    
\end{center}

Briefly, they amount to the same values used for the EMNIST experiment.
We discuss in detail the results of our experiments in the next section.

\section{Experimental results and discussion}
\label{sec:results}

In this section we detail the experiments executed on the FalconsAI model. 
All runs were executed on a AMD Ryzen 7600X CPU equipped with 32 GB of RAM, an NVIDIA RTX 3090 GPU, and a 1 TB SSD. 

\subsection{Performance}
\label{sec:performance}

\begin{figure*}[t]
    \centering
    \includegraphics[width=0.96\linewidth]{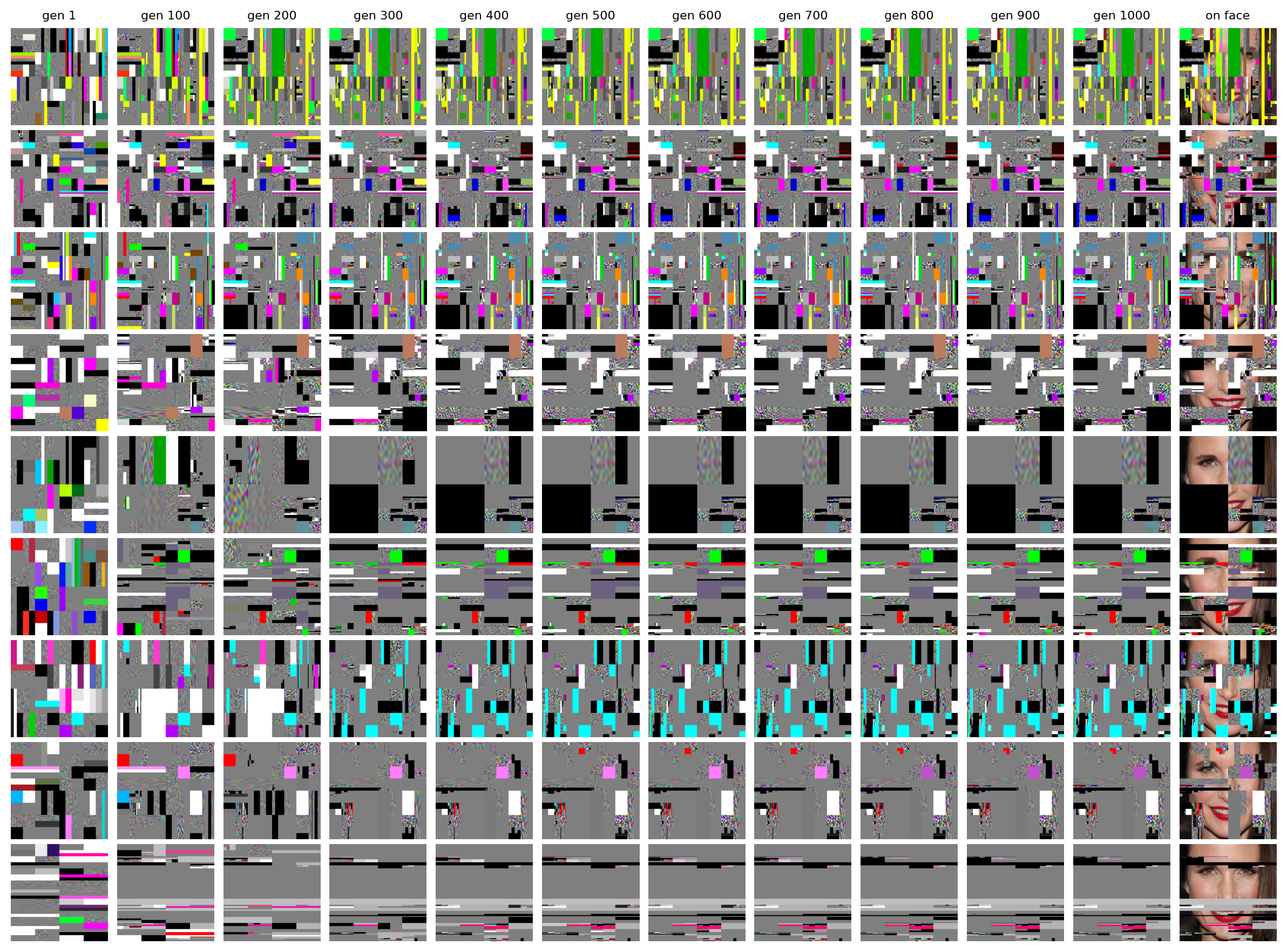}
    \caption{The best masks sampled every 100 generations and the application of the winning masks to a face.}
    \label{fig:nsfw:evolution}
\end{figure*}

\pgfplotsset{width=7cm,compat=1.18}\usepgfplotslibrary{groupplots}

\begin{figure*}[t]
\begin{tikzpicture}
\begin{groupplot}[group style={group size=3 by 3},height=3cm,width=0.38\textwidth,ytick={0.3,0.6,0.9},xtick={0,200,400,600,800}, xmin=0,xmax=1000]
    \nextgroupplot[legend columns=3, legend image post style={mark=none}, legend style={ draw=none, fill=none}, legend to name=firstplot]
        \addplot table [x=gen, y=A, col sep=comma, mark=none] {data/falconsai_eth_E1.csv};
        \addlegendentry{\texttt{nsfw}};
        \addplot table [x=gen, y expr=1-\thisrow{B}, col sep=comma, mark=none] {data/falconsai_eth_E1.csv};
        \addlegendentry{\texttt{normal}};
        \addplot table [x=gen, y=acc, col sep=comma, mark=none] {data/falconsai_eth_E1.csv};
        \addlegendentry{Overall};
        \draw[dashed, thick, gray]({rel axis cs:0,0}|-{axis cs:0,0.73}) -- ({rel axis cs:1,0}|-{axis cs:0,0.73});
    \nextgroupplot
        \addplot table [x=gen, y=A, col sep=comma, mark=none] {data/falconsai_eth_E2.csv};
        \addplot table [x=gen, y expr=1-\thisrow{B}, col sep=comma, mark=none] {data/falconsai_eth_E2.csv};
        \addplot table [x=gen, y=acc, col sep=comma, mark=none] {data/falconsai_eth_E2.csv};
        \draw[dashed, thick, gray]({rel axis cs:0,0}|-{axis cs:0,0.84}) -- ({rel axis cs:1,0}|-{axis cs:0,0.84});
    \nextgroupplot
        \addplot table [x=gen, y=A, col sep=comma, mark=none] {data/falconsai_eth_E3.csv};
        \addplot table [x=gen, y expr=1-\thisrow{B}, col sep=comma, mark=none] {data/falconsai_eth_E3.csv};
        \addplot table [x=gen, y=acc, col sep=comma, mark=none] {data/falconsai_eth_E3.csv};
        \draw[dashed, thick, gray]({rel axis cs:0,0}|-{axis cs:0,0.86}) -- ({rel axis cs:1,0}|-{axis cs:0,0.86});
    \nextgroupplot
        \addplot table [x=gen, y=A, col sep=comma, mark=none] {data/falconsai_age_A1.csv};
        \addplot table [x=gen, y expr=1-\thisrow{B}, col sep=comma, mark=none] {data/falconsai_age_A1.csv};
        \addplot table [x=gen, y=acc, col sep=comma, mark=none] {data/falconsai_age_A1.csv};
        \draw[dashed, thick, gray]({rel axis cs:0,0}|-{axis cs:0,0.65}) -- ({rel axis cs:1,0}|-{axis cs:0,0.65});
    \nextgroupplot
        \addplot table [x=gen, y=A, col sep=comma, mark=none] {data/falconsai_age_A2.csv};
        \addplot table [x=gen, y expr=1-\thisrow{B}, col sep=comma, mark=none] {data/falconsai_age_A2.csv};
        \addplot table [x=gen, y=acc, col sep=comma, mark=none] {data/falconsai_age_A2.csv};
        \draw[dashed, thick, gray]({rel axis cs:0,0}|-{axis cs:0,0.64}) -- ({rel axis cs:1,0}|-{axis cs:0,0.64});
    \nextgroupplot
        \addplot table [x=gen, y=A, col sep=comma, mark=none] {data/falconsai_age_A3.csv};
        \addplot table [x=gen, y expr=1-\thisrow{B}, col sep=comma, mark=none] {data/falconsai_age_A3.csv};
        \addplot table [x=gen, y=acc, col sep=comma, mark=none] {data/falconsai_age_A3.csv};
        \draw[dashed, thick, gray]({rel axis cs:0,0}|-{axis cs:0,0.64}) -- ({rel axis cs:1,0}|-{axis cs:0,0.64});
    \nextgroupplot
        \addplot table [x=gen, y=A, col sep=comma, mark=none] {data/falconsai_gen_G1.csv};
        \addplot table [x=gen, y expr=1-\thisrow{B}, col sep=comma, mark=none] {data/falconsai_gen_G1.csv};
        \addplot table [x=gen, y=acc, col sep=comma, mark=none] {data/falconsai_gen_G1.csv};
        \draw[dashed, thick, gray]({rel axis cs:0,0}|-{axis cs:0,0.66}) -- ({rel axis cs:1,0}|-{axis cs:0,0.66});
    \nextgroupplot
        \addplot table [x=gen, y=A, col sep=comma, mark=none] {data/falconsai_gen_G2.csv};
        \addplot table [x=gen, y expr=1-\thisrow{B}, col sep=comma, mark=none] {data/falconsai_gen_G2.csv};
        \addplot table [x=gen, y=acc, col sep=comma, mark=none] {data/falconsai_gen_G2.csv};
        \draw[dashed, thick, gray]({rel axis cs:0,0}|-{axis cs:0,0.66}) -- ({rel axis cs:1,0}|-{axis cs:0,0.66});
    \nextgroupplot
        \addplot table [x=gen, y=A, col sep=comma, mark=none] {data/falconsai_gen_G3.csv};
        \addplot table [x=gen, y expr=1-\thisrow{B}, col sep=comma, mark=none] {data/falconsai_gen_G3.csv};
        \addplot table [x=gen, y=acc, col sep=comma, mark=none] {data/falconsai_gen_G3.csv};
        \draw[dashed, thick, gray]({rel axis cs:0,0}|-{axis cs:0,0.64}) -- ({rel axis cs:1,0}|-{axis cs:0,0.64});
    \end{groupplot}
    \path (7,-5.5) node {\pgfplotslegendfromname{firstplot}};
\end{tikzpicture}
\caption{Evolution of the best individual for the nine experiments on FalconsAI.}
\label{fig:nsfwplot}
\end{figure*}
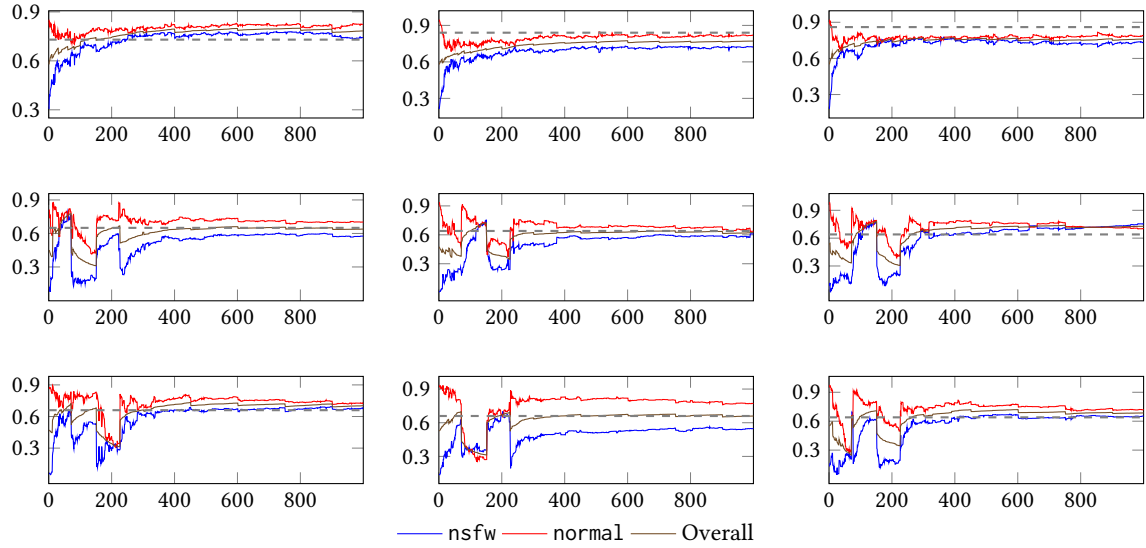

For each target task, i.e., ethnicity, age and gender, we report three distinct runs of \toolname{}.
For each run, Figure~\ref{fig:nsfw:evolution} displays the best mask sampled every 100 generations and the application of the winning mask to a face.
Similar to what we did for the EMNIST experiment (see  Figure~\ref{fig:emnistplot}), Figure~\ref{fig:nsfwplot} shows the evolution of the accuracy for the nine experiments.
For each run, we report the accuracy of the two labels \texttt{nsfw} and \texttt{normal} (blue and red lines) as well as the overall accuracy of the obtained classifier (yellow line) at training time.
Also, the dashed line indicates the accuracy at validation time.

\begin{table}[t]
    \centering
    \caption{Performance of the experiments on FalconsAI.}
    \label{tab:falconsexp}
    \setlength{\tabcolsep}{4pt}
    \begin{tblr}{
            colspec={|ll|ccc|ccc|ccc|},
            row{4,6,8,10,12,14,16,18,20,22}={gray!25},
            }
        \hline
        & & \SetCell[c=3]{c} Ethnicity & & & \SetCell[c=3]{c} Age & & & \SetCell[c=3]{c} Gender & & \\
        \cline{3-11}
        & & Run 1 & Run 2 & Run 3 & Run 1 & Run 2 & Run 3 & Run 1 & Run 2 & Run 3 \\
        \hline
        \SetCell[r=8]{c,bg=white} nsfw & TP & 158 & 195 & 211 & 155 & 175 & 184 & 188 & 174 & 125 \\
        & TN & 205 & 224 & 219 & 172 & 146 & 137 & 142 & 157 & 193 \\
        & FP & 45 & 26 & 31 & 78 & 104 & 113 & 108 & 93 & 57 \\
        & FN & 92 & 55 & 39 & 95 & 75 & 66 & 62 & 76 & 125 \\
        & Acc & 0.73 & 0.84 & \underline{0.86} & \underline{0.65} & 0.64 & 0.64 & \underline{0.66} & \underline{0.66} & 0.64 \\
        & Prec & 0.78 & \underline{0.88} & 0.87 & \underline{0.67} & 0.63 & 0.62 & 0.64 & 0.65 & \underline{0.69} \\
        & Rec & 0.63 & 0.78 & \underline{0.84} & 0.62 & 0.70 & \underline{0.74} & \underline{0.75} & 0.70 & 0.50 \\
        & F1 & 0.70 & 0.83 & \underline{0.86} & 0.64 & 0.66 & \underline{0.67} & \underline{0.69} & 0.67 & 0.58 \\
        \hline
        \SetCell[r=8]{c,bg=white} normal & TP & 205 & 224 & 219 & 172 & 146 & 137 & 142 & 157 & 193 \\
        & TN & 158 & 195 & 211 & 155 & 175 & 184 & 188 & 174 & 125 \\
        & FP & 92 & 55 & 39 & 95 & 75 & 66 & 62 & 76 & 125 \\
        & FN & 45 & 26 & 31 & 78 & 104 & 113 & 108 & 93 & 57 \\
        & Acc & 0.73 & 0.84 & \underline{0.86} & \underline{0.65} & 0.64 & 0.64 & \underline{0.66} & \underline{0.66} & 0.64 \\
        & Prec & 0.69 & 0.80 & \underline{0.85} & 0.64 & 0.66 & \underline{0.67} & \underline{0.70} & 0.67 & 0.61 \\
        & Rec & 0.82 & \underline{0.90} & 0.88 & \underline{0.69} & 0.58 & 0.55 & 0.57 & 0.63 & \underline{0.77} \\
        & F1 & 0.75 & 0.85 & \underline{0.86} & \underline{0.67} & 0.62 & 0.60 & 0.63 & 0.65 & \underline{0.68} \\
        \hline
        \SetCell[r=3]{c,bg=white} global & True & 363 & 419 & 430 & 327 & 321 & 321 & 330 & 331 & 318 \\
        & False & 137 & 81 & 70 & 173 & 179 & 179 & 170 & 169 & 182 \\
        & Acc & 0.73 & 0.84 & \textbf{0.86} & \textbf{0.65} & 0.64 & 0.64 & \textbf{0.66} & \textbf{0.66} & 0.64 \\
        \hline
    \end{tblr}
\end{table}

In Table~\ref{tab:falconsexp} we give the performance details of each task, i.e., ethnicity, age and gender.
For each task, we carried out three experiments, reported in columns Run 1-3.
The rows are grouped into three blocks.
The first two blocks report the confusion matrix values and per-class metrics for \texttt{nsfw} and \texttt{normal}, respectively.
The third block (global) reports the values for the resulting binary classifier.
In this case, we just need to report correct predictions (\emph{T}) and wrong ones (\emph{F}), together with the overall accuracy of the resulting classifier.
Bold values mark the best global result per task across the three runs; underlined values denote the best result within the same per-class block (\texttt{nsfw} or \texttt{normal}) across the three runs of the same task.

In terms of computation time, the longest execution took 5 hours and 15 minutes, while the shortest one completed in 4 hours and 48 minutes.
On average, the fuzzing algorithm requires $\sim$18 seconds per generation.

\subsection{Possible countermeasures}
\label{sec:countermeasures}

For a detailed survey on defenses against backdoors, we refer the interested reader to~\cite{9743317},~\cite{zhang2024backdoor} and~\cite{10.1145/3547330}.
Since our black-box attacker model for installing a backdoor is novel, no specific defense mechanisms exist in the literature. 
In fact, most approaches target backdoors activated before the model deployment, e.g., due to data or model poisoning.
Nevertheless, some techniques could counter our methodology.
In particular, here we distinguish between \emph{countermeasures}, which counter the vulnerability, and \emph{mitigations}, which reduce its exploitability. The main countermeasures against \toolname{} are the following.
\begin{enumerate}
    \item \emph{Fairness auditing and adversarial debiasing}. 
    Fairness audits identify demographic asymmetries in a model's predictions, and adversarial debiasing attempts to remove them by retraining with fairness-aware objectives. Their effectiveness against \toolname{} depends on the source of the latent bias being surfaced: if the bias originates from training-data imbalance, then debiasing the training pipeline directly shrinks the attack surface; on the contrary, if the bias is a structural property of the primary task itself, the technique is largely ineffective. Our EMNIST case study illustrates the latter regime: \toolname{} activates the backdoor on a digit classifier trained on MNIST, which carries no demographic bias.

    \item \emph{Invariance testing}. 
    Before deployment, the model owner can systematically probe the classifier with structured input transformations and measure whether its outputs exhibit stable directional shifts across
    demographic or other latent biases. Invariance testing is a form of robustness testing in ML workflows~\cite{9898127}, and one of its uses is precisely to surface spurious correlations or biases that the model has internalized. Persistent directional shifts under such probes signal exactly the kind of latent bias that \toolname{} would later amplify at inference time. While this approach does not by itself remove the bias, it gives the deployer a diagnostic tool to decide whether the model is suitable for deployment, or whether further interventions (debiasing, retraining, or replacement) are advisable. We regard this as a plausibly effective countermeasure rather than one we have empirically tested.
    
    \item \emph{Output calibration}.
    Post-processing calibration techniques such as temperature scaling~\cite{10.5555/3305381.3305518} address \emph{confidence calibration} -- the problem of producing probability estimates representative of the true correctness likelihood -- by rescaling a model's predicted probabilities to mitigate overconfidence, without modifying the underlying classification decisions. We note that the fitness function driving our genetic search (Equation~\ref{eq:fitness}) consumes score-based feedback to discriminate the two target groups, so reducing the variance of that probability distribution would, in principle, weaken the signal the wrapper exploits. We did not evaluate \toolname{} on a calibrated victim, so the argument is only suggestive: calibration may attenuate the margin through which the latent bias is observed, but it does not remove the bias itself.

 \item \emph{Saliency-based defenses}.
    In perception tasks such as face recognition or NSFW detection, saliency analysis (e.g., Grad-CAM~\cite{gradcam2017} and integrated gradients)~\cite{pmlr-v70-sundararajan17a} reveals which image regions drive a model's decision and has been used both for interpretability and for targeted perturbations.
    On the defensive side, SentiNet~\cite{sentinet2020} leverages Grad-CAM to localize trigger regions in individual inputs and flag backdoor-driven predictions; this saliency-based detection operates during inference but assumes a spatially concentrated trigger, for instance, a property absent from the distributed, programmatic transformations evolved by \toolname{}.
\end{enumerate}

In terms of operational mitigations, two further measures can raise the cost of mounting \toolname{} without addressing its
root cause.
\begin{enumerate}
    \item \emph{Query monitoring}.
    Defenses such as Blacklight~\cite{li2022blacklight} detect adversarial query patterns by hashing input similarity across the inference history and flagging clusters of near-duplicate queries, on the assumption that crafting an adversarial input typically requires many small perturbations of the same starting point. The evolution phase of \toolname{} issues on the order of $10^7$~queries (Section~\ref{sec:nsfw}), many of which are derived from the same underlying mask population and therefore exhibit measurable input similarity, which ideally matches the patterns that such a defense targets. While we did not run \toolname{} against a query-monitoring deployment, rate-limiting or per-account quotas would plausibly raise the discovery cost substantially.

    \item \emph{Randomized input pre-processing}.
    The deployer can insert non-deterministic transformations between their public API and the underlying classifier (e.g.,~random shifts, JPEG re-encoding at random quality, light photometric jitter), so that two identical queries are processed through slightly different pipelines.
    This breaks the spatial assumptions of fixed evolved masks and is the deployment-time analogue of the purification defenses surveyed in~\cite{10.1145/3547330}.
    To probe \toolname{}'s resilience to this class of mitigation, we evaluate the selected single mask under four deterministic preprocessing perturbations: a translation of the cropped face by a few pixels (\emph{Shift}), a Gaussian blur (\emph{Blur}), a JPEG re-encoding at moderate quality (\emph{JPEG}), and a global brightness perturbation (\emph{Bright}); the \emph{Base} column reports the same evaluation when the masked image is fed to the classifier as-is, with no additional transformation.
    Table~\ref{tab:nsfw-robustness} reports the resulting balanced accuracy across the three target tasks; the values are computed on the development split used for mask selection, so the Base column may differ by a few decimals from the corresponding test-split accuracies of Table~\ref{tab:falconsexp}.
    The worst-case drop is $2.1\%$ from Base, observed under Shift on the age and gender tasks, while all other preset/task pairs degrade by at most $1.7\%$; on ethnicity, Blur and JPEG are essentially indistinguishable from Base.
    The mask encodes coarse, region-level color and brightness offsets rather than high-frequency adversarial patterns, and is therefore largely insensitive to mild geometric, photometric, and compression-based perturbations.
    Lightweight randomization alone is unlikely to neutralize the attack; stronger or adversarially-tuned randomization schedules remain to be evaluated.
\end{enumerate}

\begin{table}[t]
\centering
\begin{minipage}[t]{0.49\linewidth}
    \centering
    \caption{\toolname{} robustness test against perturbations.}
    \label{tab:nsfw-robustness}
    \setlength{\tabcolsep}{4pt}
    \begin{tabular}{|c|c|cccc|}
    \hline
    Task & Base & Shift & Blur & JPEG & Bright \\
    \hline
    E & 0.849 & 0.832 & 0.849 & 0.846 & 0.838 \\
    A & 0.664 & 0.643 & 0.664 & 0.662 & 0.653 \\
    G & 0.696 & 0.675 & 0.698 & 0.697 & 0.693 \\
    \hline
    \end{tabular}
\end{minipage}\hfill
\begin{minipage}[t]{0.49\linewidth}
    \centering
    \caption{Performance improvement with WMA.}
    \label{tab:nsfw-tasks}
    \setlength{\tabcolsep}{4pt}
    \begin{tabular}{|c|ccc|ccc|}
    \hline
    Task & 1 & 2 & 3 & WMA & Gain & Max. \\
    \hline
    E & 0.73 & 0.84 & 0.86 & 0.87 & $+0.01$ & 0.85 \\
    A & 0.65 & 0.64 & 0.64 & 0.68 & $+0.03$ & 0.70 \\
    G & 0.66 & 0.66 & 0.64 & 0.69 & $+0.03$ & 0.75 \\
    \hline
    \end{tabular}
\end{minipage}
\end{table}

\subsection{Further considerations}
\label{sec:discussion}

Here we put forward some further aspects related to our technique as well as emerging phenomena that we observed during our experiments.

\paragraph*{Performance improvement}
Although our experiments confirm the effectiveness of \toolname in enabling backdoors in the victim classifier, we observed that the obtained $T'$ classifier's accuracy may significantly vary.
For instance, in Table~\ref{tab:falconsexp} we observe that, under similar settings, we achieved 86\% accuracy for the ethnicity task, but only 0.65 for the age task.

This result may be due to various factors.
We may consider some tasks to be intrinsically more complex than others.
For instance, telling whether a young person in a picture is 24 or 26 is even hard for human beings.
Other factors may be related to the internals of the target classifiers that we intentionally neglect in our black-box attack.
In general, further research is needed to find techniques ensuring better accuracy results.

Nevertheless, as discussed above, the attacker can obtain a reasonably accurate mask in a few hours of computations.
Thus, rather than refining such a mask by extending the fuzzing generations, she may opt for implementing a \emph{weighted majority algorithm} (WMA) \cite{LITTLESTONE1994212}.
Such a technique permits combining multiple classifiers through a weighted voting system, so that the global accuracy surpasses the individual one.

Table~\ref{tab:nsfw-tasks} shows the results obtained by implementing WMA with the masks of Section~\ref{sec:nsfw}.
For each task, we report the accuracy of the three masks, the accuracy of the WMA classifier and the accuracy gain with respect to the best mask.
For instance, WMA resulted in an 87\% accuracy for the ethnicity task (first row), corresponding to a $+1\%$.
In general, if the attacker has many masks to combine, the performance should improve further.
Also, WMA performs better when the individual classifiers are statistically independent.
In the right column of Table~\ref{tab:nsfw-tasks}, we indicate the maximum accuracy theoretically achievable if the three masks used by WMA were statistically independent, computed following Kuncheva~\cite{kuncheva2004combining}.
Currently, \toolname{} does not guarantee this property, but this direction can be investigated as future work.

Another possible improvement is related to the mask generation costs.
In particular, if querying the victim classifier has a cost, we may want to estimate the number of such queries.
Each fuzzing round evaluates the full population of $\pi$ MET on a balanced training subset of $n \cdot \omega$ samples.
For instance, under the settings of Section~\ref{sec:nsfw:discovery} ($\pi = 32$, $n=2$, $\omega = 400$, and $1000$ generations), the number of queries is $32 \cdot 2 \cdot 400 \cdot 1000 \approx 2.6 \times 10^7$ in the worst case.
Although this cost is paid only once for each mask, it is worth investigating optimization techniques for reducing it.
This also accounts as future work.



\paragraph*{Inferring model biases}
As discussed above, our technique works by mapping the backdoor task into an existing model bias.
Although this allows us to carry out our attack, the same approach may be adapted to search and test model biases.
In particular, consider this statement: FalconsAI NSFW training data has a bias toward a specific ethnicity.
For instance, since the model was trained with an unbalanced dataset containing a majority of white people in unsafe contexts.
Clearly, testing whether the statement above is true or false would be interesting.

In principle, if the statement is false, when generating a mask for the ethnicity task of Section~\ref{sec:nsfw}, \toolname should map white or black faces to \texttt{nsfw} with the same probability.
That would indicate that the backdoor task has no correlation with the actual model behavior.
However, by running \toolname{} on FalconsAI for the ethnicity task 12 times, we noticed that in all the cases, the generated masks map ``white'' into \texttt{nsfw}. 
Although this is not enough to conclude that the bias discussed above actually occurs in FalconsAI, it suggests that further investigation could be relevant.
In particular, we are interested in studying whether \toolname{} can be turned into a bias testing methodology.

\section{Conclusion}
\label{sec:conclusion}

This paper introduced Backdoor-as-a-Service (\toolname{}), a novel backdoor attack paradigm operating in full black-box mode.
To the best of our knowledge, our proposal is the first one requiring no access to training data or model parameters.
We even confirmed the feasibility by applying our tool to FalconsAI, a state-of-the-art NSFW classification product.
According to our investigations, no countermeasures exist that can completely neutralize our attack.
As a consequence, many online classifiers may be exposed to our attack scenario.
For this reason we released \toolname{} as an open source project that developers can use to test their systems.

\bibliographystyle{ACM-Reference-Format}
\bibliography{biblio}
\end{document}